\documentclass[preprint,aps,prb,amsmath]{revtex4}
\usepackage{epsfig}
\usepackage{amsmath}
\usepackage{amsfonts}
\usepackage{graphicx}
\usepackage{dcolumn}%
\usepackage{bm}%
\usepackage{amssymb}
\usepackage{color}

\def\be{\begin{equation}}
\def\ee{\end{equation}}

\def\r{{\bm{r}}}
\def\lan{\langle}
\def\ran{\rangle}

\begin{document}

\title{Emergence of complex behaviour in gelling systems starting 
from simple behaviour of single clusters}
\author{A. Fierro$^{a,b}$, T. Abete $^{a,b}$, and A. Coniglio$^{a,b,c}$}
\affiliation{${}^a$ INFM-CNR Coherentia\\
${}^b$
Dipartimento di Scienze Fisiche, Universit\`a di
Napoli ``Federico II'',\\ Complesso Universitario di Monte
Sant'Angelo, via Cintia 80126 Napoli, Italy
\\
${}^c$ INFN Udr di Napoli}

\begin{abstract}

A theoretical and numerically study of dynamical properties in the
sol-gel transition is presented. In particular, 
the complex phenomenology observed experimentally and numerically in gelling
systems is reproduced in the framework of percolation 
theory, under simple assumptions on the relaxation of single clusters. 
By neglecting the correlation between particles belonging to
different clusters, the quantities of interest (such as
the self Intermediate Scattering Function, the dynamical susceptibility,
the Van-Hove function, and the non-Gaussian parameter) are 
written as superposition of those due to single clusters.
Connection between these behaviours and the critical exponents of
percolation are given.
The theoretical predictions are checked in a 
model for permanent gels, where bonds between monomers are described by a 
FENE potential. The data obtained in the numerical simulations
are in good agreement with the analytical predictions.

\end{abstract}

\maketitle

\section{Introduction}
\label{sect1}

The gelation transition transforms a viscous liquid (sol) into an 
elastic disordered solid (gel). In general this process is due to
the formation of a macroscopic molecule, due to the bonding
of multifunctional monomers in solution, which makes the
system able to bear stress. The extent of the gelation 
process may be measured by the monomer volume fraction $\phi$,
defined as $\phi=N V_m/V$, where $N$ is the number of monomers,
$V_m$ is the single monomer volume and $V$ is the total system volume.
On the static point of view, the sol-gel transition 
was interpreted \cite{flo,deg} in terms of the appearance of 
a percolating cluster of monomers linked by bonds \cite{stauffer}, and 
experimental measurements of the geometric properties of gels
have confirmed this correspondence (for a review see Stauffer et 
al.\cite{adconst} and references therein). 

Complex dynamical behaviours are observed in gelling systems already in the sol
phase. For example, 
light scattering measurements show
non-exponential decay of the intermediate scattering function,
$F(k,t)$, in both permanent \cite{martin} 
and thermoreversible physical gels \cite{ikkai,ren}.
In particular power laws are observed at 
intermediate times, followed, at long times, by stretched exponential decays,
$\exp(-(t/\tau)^\beta)$, with $0<\beta<1$.
In the gel phase, where ergodicity is broken, only the power law decay survives.
Usually the onset of stretched exponential decays (present also in other
complex systems, as  spin glasses and  glassy systems)
is associated to the widening of 
relaxation times, which in gelling systems is due to the presence 
of a broad cluster size distribution close 
to the gelation threshold. However general predictions which connect this kind
of relaxation to percolation theory are not easily feasible.

In this paper, assuming ``simple'' behaviours for the relaxation of clusters 
with given size, we show how the ``complex'' phenomenology of the relaxation in
permanent gels may be obtained from the superposition of the  behaviours of 
clusters with different sizes.
In particular we are able to predict the  
behaviours of the 
self Intermediate Scattering Function, of the dynamical susceptibility,
of the Van-Hove function, and of the non-Gaussian parameter, and to connect
these behaviours to the cluster size distribution and the critical exponents of
percolation.
Then we check the theoretical predictions in a specific model for
permanent gels,  studied using Molecular Dynamics simulations.

The paper is organized as follows.
In Sect.\ref{sect2} the analytical results are briefly summarized, and
in Sect.\ref{sect3} they are compared with the data 
obtained by Molecular Dynamics simulations of a model for
permanent gels, where bonds between monomers are described 
by a FENE potential \cite{FENEdum,FENE,tiziana_prl}. In Sect.\ref{sect4} 
concluding 
remarks are
discussed. Finally, in \ref{appendix1}, \ref{appendix2} and \ref{appendix3}
the calculations are presented in details.

\section{Connection between static and dynamic properties}
\label{sect2}

In this section we summarize our calculations, which will be show in details in
appendices. 
We consider a system of randomly distributed monomers with a fixed volume 
fraction, $\phi$. At time $t=0$ permanent bonds are
introduced at random between monomers at a distance $r<R$, where $R$ is 
suitably chosen.  For a particular model see the FENE model \cite{FENEdum,FENE,
tiziana_prl} (Sect. \ref{sect4}), however the following arguments are 
independent on the details of the model. 

Following the percolation approach \cite{flo,deg}, we identify the gel phase as
the state where a percolating cluster is present.
We denote by $\phi_c$ the volume
fraction of the percolation threshold.
In our calculations we use some results from percolation theory 
\cite{stauffer}: 
In particular, in the sol phase, near the threshold,
the cluster size distribution is given by $n(s)\propto s^{-\tau} e^{-s/s^*}$ 
(where $s^*$ is
a cut-off value given by $\xi^{D_f}$, $D_f$ is the fractal dimension,
and $\xi$ is the connectedness
length which diverges at the threshold with the exponent $\nu$);
in the gel phase, near the threshold,
$sn(s)$ is put equal to 
$P_\infty\delta_{s,s_{max}}+C s^{-\tau+1}e^{-(s/s^*)^{(d-1)/d}}$, 
where $P_\infty$ is the density of particles in
the percolating cluster of mass $s_{max}$, $d$ is the spatial dimension 
and $C$ is a constant.
Moreover we assume that the relaxation time of clusters
increases as a power law of the size \cite{theoryandexperimets}, 
$\tau(s)\propto s^{x}$.
With these assumptions, in the hypothesis of simple behaviour for single 
cluster (exponential
relaxation, simple diffusion, etc.), we obtain all the complex 
phenomenology observed experimentally and numerically
near the threshold in gelling systems.
In particular we are able
to predict the  
behaviours of the self Intermediate
Scattering Function, of the dynamical susceptibility,
of the Van-Hove function, and of the non-Gaussian parameter.

\subsection{Self Intermediate Scattering Functions}
We first consider the self Intermediate Scattering Functions (ISF):
\begin{equation}
F_{self}(k,t)=\left[\lan \Phi_{self}(k,t)\ran\right]
\end{equation}
where $\langle \dots \rangle$ is the thermal average over a fixed
bond configuration, $\left[\dots\right]$ is the average over
independent bond configurations of the system, 
\begin{equation}
\Phi_{self}(k,t)=\frac{1}{N}\sum_{i=1}^N e^{i\vec{k}
\cdot(\vec{r}_i(t)- \vec{r}_i(0))},
\end{equation}
and $N$ is the number of particles.  
In the following we fix the wave vector $k=k_{min}$ and $k_{min}=2\pi/L$, with 
$L$ the linear system dimension. 

In terms of the  contributions due to different clusters, 
$F_{self}(k_{min},t)$ can be written
as 
\begin{equation}
F_{self}(k_{min},t)=\left[ \sum_s s n(s) f_s(k_{min},t) \right],
\label{eq5}
\end{equation}
where $n(s)$ is the cluster size distribution ($N n(s)$ gives the number of
clusters of size $s$) 
 and $f_s(k,t)=\overline{\lan  f_{C_s}(k,t)\ran}$, where
$f_{C_s}(k,t)=\frac{1}{s}\sum_{i\in C_s}
e^{i\vec{k}\cdot(\vec{r}_i(t)-\vec{r}_i(0))}$ is the self ISF limited to a
given cluster $C_s$ of size $s$, and 
$\overline{{{\dots^{}}^{}}^{}}$ is the average over all clusters of 
given size $s$.

By replacing the sum with the integral, the self ISF for a given bond 
configuration becomes:
\begin{eqnarray}
F_{self}(k_{min},t)\sim\int ds~sn(s) f_s(k_{min},t).
\label{integral}
\end{eqnarray}
By assuming
\begin{equation}
f_s(k_{min},t) \simeq e^{-t/\tau(s)},
\label{eq:exp_simple}
\end{equation}
the integral in Eq.(\ref{integral}) 
gives in the thermodynamic limit 
the following predictions for the time dependence of
$F_{self}(k_{min},t)$ in a permanent gel:
\begin{enumerate}
\item[(i)] At the gelation threshold ($\phi=\phi_c$)
\begin{equation}
F_{self}(k_{min},t)\propto~t^{-z}\Gamma(z), \label{conto_soglia}
\end{equation}
where $\Gamma(z)\equiv\int d\sigma~\sigma^{z-1}\exp(-\sigma)$ is the
$\Gamma$-function with $z=(\tau-2)/x$.

\item[(ii)] In the sol phase ($\phi<\phi_c$)
\begin{equation}
F_{self}(k_{min},t)\propto~t^{-c_1} e^{-(t/\tau_\alpha)^\beta},
\label{conto4}
\end{equation}
where $\beta=1/(x+1)$, $c_1=\beta(\tau-3/2)$, and
$\tau_\alpha\propto~(\phi_c-\phi)^{-f}$, and
$f=x D_f \nu$. This approximated form, obtained in
the long time limit, coincides with that
suggested by Ogielski\cite{ogielski} as fitting
function for the time dependent order parameter in spin glasses, and
it is in agreement with experimental \cite{martin} and numerical \cite{cubetti}
findings in gelling systems.

\item[(iii)] In the gel phase  ($\phi>\phi_c$)
\begin{equation}
F_{self}(k_{min},t) \simeq P_\infty+At^{-c_1} e^{-(t/\tau_\alpha)^\beta},
\label{conto_gel}
\end{equation}
\end{enumerate}
where $\beta=1/(x+1)$, and $c_1=\beta(\tau-3/2)$ are the same exponents 
obtained in the sol phase. The plateau value, $P_\infty$, gives the density
of {\em localized} particles \cite{goldbart}. 
Clearly the main contribution comes from
localized particles of the percolating cluster, however a small contribution
may be due to particles trapped inside it.

The calculations are shown in details in \ref{appendix1}.

Our findings given in Eq.s (\ref{conto_soglia}), (\ref{conto4}) and
Eq.(\ref{conto_gel})
are in agreement with the theoretical predictions obtained in Ref.\cite{zippelius}
in the Rouse and Zimm models for randomly cross-linked monomers, where
$x=1$ and $x=1/2$ respectively.
Similar calculations are also done in Ref.\cite{sontolongo} in a different
context.

\subsection{Fluctuations of the self ISF}
In Ref. \cite{tiziana_prl} it was studied the dynamical susceptibility, 
defined as the fluctuations of
the self ISF:
\begin{equation}
\chi_4(k,t)=N\left[\rule{0pt}{10pt}\lan |\Phi_{self}(k,t)|^2\ran-\lan
\Phi_{self}(k,t)\ran^2\right].
\label{chi4}
\end{equation}
In particular it was shown that, in the sol phase, in the limit of 
$t\to\infty$ and $k\to 0$, $\chi_4(k,t)$ coincides with the mean cluster size, 
$S=\sum_s
s^2 n(s)$: \be \lim_{k\to 0}\lim_{t\to\infty} \chi_4(k,t)=S, \ee
which diverges at the threshold \cite{tiziana_prl} with the exponent $\gamma$. 
Here we are interested in
the time dependence of the dynamical susceptibility approaching the
asymptotic value.

We neglect the contributions due to disconnected particles at each time $t$.
In this way we can write $\chi_4(k,t)$ as a superposition of the contributions
due to different clusters:
\be
\chi_4(k,t)\simeq \left[
\sum_s s^2 n(s) \overline{\lan f_{C_s}(k,t)f^*_{C_s}(k,t)\ran-
\lan f_{C_s}(k,t)\ran\lan f^*_{C_s}(k,t)\ran}  \right],
\label{eq10}
\ee
where again $f_{C_s}(k,t)=\frac{1}{s}\sum_{i\in C_s}
e^{i\vec{k}\cdot(\vec{r}_i(t)-\vec{r}_i(0))}$ is the self ISF limited to a
given cluster $C_s$ of size $s$,
$\overline{{{\dots^{}}^{}}^{}}$ is the average over all clusters of
given size $s$,
$\lan\dots\ran$ is the thermal average over a fixed bond
configuration, and $\left[\dots\right]$ is the average over independent bond
configurations. The term $\lan f_{C_s}(k,t)f^*_{C_s}(k,t)\ran$
in Eq.(\ref{eq10}) can be written as
\begin{eqnarray}
\lan f_{C_s}(k,t)f^*_{C_s}(k,t)\ran&&=\frac{1}{s^2}\sum_{i,j\in C_s}\lan
e^{i\vec{k}\cdot(\vec{r}_i(t)-\vec{r}_j(t))}
e^{-i\vec{k}\cdot(\vec{r}_i(0)-\vec{r}_j(0))}\ran \nonumber \\
&&=
\frac{1}{s^2}\sum_{i,j\in C_s}
\lan e^{i\vec{k}\cdot\vec{\Delta}_{ij}(t)}\ran,
\end{eqnarray}
where we have put $\vec{\Delta}_{ij}(t)\equiv
(\vec{r}_i(t)-\vec{r}_j(t))-(\vec{r}_i(0)-\vec{r}_j(0))$.
For connected particles $i$ and $j$, $|\vec{\Delta}_{ij}(t)|$ is finite, and 
in the low wave vector limit where $|\vec{\Delta}_{ij}(t)|\ll 1/k$
 we can assume $e^{i\vec{k}\cdot\vec{\Delta}_{ij}(t)}\simeq 1$. 
Then, by supposing that $\overline{\lan f_{C_s}(k,t)\ran 
\lan f^*_{C_s}(k,t)\ran} = \overline{\lan f_{C_s}(k,t) \ran}^2$,
in the zero wave vector limit the dynamical susceptibility for a given bond 
configuration can be written as:
\be
\lim_{k\to 0}\chi_4(k,t)\simeq \lim_{k\to 0}\sum_s s^2 n(s )\left(1-
\lan f_s(k,t)\ran^2\right).
\label{eq16}
\ee

From this equation, using Eq.(\ref{eq:exp_simple}), 
it is direct to see that the dynamical
susceptibility goes from zero (for $t=0$) to the mean cluster size
$S$ (in the $t\to\infty$ limit), since the self ISF of clusters of
given size, in the sol phase, goes from $1$ (for $t=0$) to zero (in the
$t\to\infty$ limit).

As in the previous section 
we can evaluate 
$\lim_{k\to 0}\chi_4(k,t)$ in the sol phase, $\phi<\phi_c$.
We find that, for time long enough, $\chi_4(k,t)$ approaches the asymptotic 
value 
in the following way:
\begin{equation}
\lim_{k\to 0}\chi_4(k,t)\simeq~S\cdot\left(1-At^{c_2} e^{-(2
t/\tau_\alpha)^\beta}\right), \label{conto6}
\end{equation}
where $\beta=1/(x+1)$ and $c_2=\beta(5/2-\tau)$.
The exponent $\beta$ is exactly the same which appears in Eq.(\ref{conto4}) 
for the decay
to zero of the self ISF, the relaxation time in the stretched exponential
function is given by $\tau_\alpha/2$, and finally the
power law has a positive exponent $c_2$ different from the exponent
$c_1$ which appears in Eq.(\ref{conto4}).

The calculations are shown in details in \ref{appendix2}.

\subsection{Self part of the Van-Hove function}
The self part of the Van-Hove function \cite{hansen} is given by:
\be
G_{self}(r,t)=\frac{1}{N}\left[\langle\sum_{i=1}^N\delta
(r-|\r_i(t)-\r_i(0)|) \rangle\right].
\label{gself0}
\ee
If the motion of particles is diffusive
with a diffusion coefficient D,
$G_{self}(r,t)=(1/4\pi Dt)^{3/2}\exp(-r^2/4Dt),$ where $r$ is the
distance traveled by a particle in a time $t$.
Deviations from the Gaussian distribution were observed in different glassy 
and gelling systems \cite{stariolo, tiziana_pre}. 
In fact the van-Hove function
seems fitted by a Gaussian only for short distances, instead, for long 
distances, it is well fitted by an exponential tail that extends to 
larger distances for increasing times.

The deviation from the Gaussian distribution
indicates that some particles move faster than
others, due to the presence of heterogeneities. 
In permanent gels, 
heterogeneities coincide with clusters of particles connected
by bonds \cite{tiziana_prl}. As matter of fact particles belonging to
different clusters have a different diffusion coefficient depending
on the cluster size. As a consequence it has been suggested \cite{tiziana_pre}
that, in the sol phase and in the diffusive regime 
(i.e. in the long time limit), 
$G_{self}(r,t)$  is given by a superposition of Gaussians
\begin{equation}
G_{self}(r,t)=\left[\sum_s \frac{s n(s)}{(4\pi
D(s)t)^{3/2}}\exp\left(-\frac{r^2}{4D(s)t}\right)\right],
\label{self-van-hove}
\end{equation}
where $D(s)$ is the diffusion coefficient of clusters of size $s$
and $n(s)$ is the cluster size distribution. 

By assuming  
$\tau^{-1}(s)\propto D(s)=a s^{-x}$, and by replacing the sum with the 
integral in Eq.(\ref{self-van-hove}),
predictions can be given for the dependence of $G_{self}(r,t)$
on $r$ and $t$. We find, in the limit $r^2\gg at$: 
\begin{equation}
t^{3/2}G_{self}(r,t)\propto    
\left(-A+\frac{1}{s^*}+\frac{xr^2}{4at}\right)^{-1}
\exp\left[-\frac{r^2}{4at}\right],
\label{eq17} 
\end{equation}
where $A\equiv 1-\tau+3x/2$.

The calculations are shown in details in \ref{appendix3}.

\subsection{Non-Gaussian parameter}

The non-Gaussian parameter is defined as
\cite{Rahman}:
\begin{equation}
\label{eq:alfa} \alpha_2(t)=\left[\frac{3 \lan\Delta r^4(t)\ran} {5 (
\lan \Delta r^2(t)\ran) ^2}\right]-1,
\end{equation}
where $\lan\Delta r^2(t)\ran=\frac{1}{N}\sum_{i=1}^N\langle
|\vec{r}_i(t)- \vec{r}_i(0)|^2\rangle$, and
$\lan\Delta r^4(t)\ran=\frac{1}{N}\sum_{i=1}^N\langle
|\vec{r}_i(t)- \vec{r}_i(0)|^4\rangle$. It is easy to show \cite{Rahman}
that $\alpha_2(t)$
is zero if the probability distribution of the particle
displacements is Gaussian.

Using Eq.(\ref{self-van-hove}), in permanent gels
the non-Gaussian parameter is expected 
to tend in the long time limit to a plateau, whose value is given by
\begin{equation}
\alpha^{as}_2=\left[\frac{\sum_s s n(s) D^2(s)}{(\sum_s s n(s)
D(s))^2}\right]-1= \left[\frac{\overline{ D ^2} -\overline{D}
^2}{\overline{D} ^2}\right], \label{fluttD}
\end{equation}
where, for each bond configuration, $\overline{\dots^{}}$  is the
average over the cluster distribution. 
From this relation it appears clear that the deviation from gaussianity 
($\alpha_2\ne 0$) is due to the fluctuations of the diffusion coefficient, 
which in turns is related to the presence of dynamical heterogeneities,
i.e. groups of particles with different diffusion coefficient.

\section{FENE model for permanent gels}
\label{sect3}
In this section we check the theoretical predictions obtained in
Sect. \ref{sect2} in the FENE model for permanent gels.
We consider a $d=3$ system of $N$ particles interacting
with a soft potential given by Weeks-Chandler-Andersen (WCA)
potential \cite{chandler}:
\begin{equation}
U_{ij}^{WCA}=\left\{ \begin{array}{ll}
4\epsilon[(\sigma/r_{ij})^{12}-(\sigma/r_{ij})^6+\frac{1}{4}], & r_{ij}<2^{1/6}\sigma \\
0, & r_{ij}\ge2^{1/6}\sigma \end{array} \right.
\end{equation}
where $r_{ij}$ is the distance between the particles $i$ and $j$.

After the equilibration, at a given
time $t=0$ particles distant less than $R_0$ are
permanently linked by adding an attractive potential:
\begin{equation}
U_{ij}^{FENE}=\left\{ \begin{array}{ll}
-0.5 k_0 R_0^2 \ln[1-(r_{ij}/R_0)^2], & r_{ij}< R_0\\
\infty, & r_{ij}\ge R_0 \end{array} \right.
\end{equation}
representing a finitely extendable nonlinear elastic (FENE). The
FENE potential was firstly introduced by Warner\cite{FENEdum} and is
widely used to study linear polymers \cite{FENE}. We choose
$k_0=30\epsilon/\sigma^2$ and $R_0=1.5\sigma$ as usual\cite{FENE}
in order to avoid any bond crossing and to use an integration time
step $\Delta t$ not too small. 

We have performed Molecular Dynamics simulations of this model
\cite{tiziana_prl}: The equations of motion were solved in the
canonical ensemble (with a Nos\'e-Hoover thermostat) using the
velocity-Verlet algorithm \cite{Nose-Hoover} with a time step
$\Delta t=0.001\delta\tau$, where
$\delta\tau=\sigma(m/\epsilon)^{1/2}$ is the standard unit time for
a Lennard-Jones fluid and $m$ is the mass of particle. We use
reduced units where the unit length is $\sigma$, the unit energy is
$\epsilon$ and the Boltzmann constant $k_B$ is set equal to $1$. We
choose periodic boundary conditions, and average all the investigated
quantities over $32$ independent configurations of the system.  The
temperature is fixed at $T=2$ and the volume fraction
$\phi=\pi\sigma^3N/6L^3$ (where $L$ is the linear size of the
simulation box in units of $\sigma$) is varied from $\phi=0.06$ to
$\phi=0.12$.

Using the percolation approach, we identify the
gel phase as a state where a percolating cluster is present
\cite{flo,deg}. With a finite size scaling analysis
\cite{tiziana_prl} we obtain that this transition is in the
universality class of random percolation.
In particular, we obtain that the cluster size distribution,
$n(s)\sim s^{-\tau}$ at the gelation threshold $\phi_c=0.09\pm0.01$, 
with $\tau=2.1\pm0.2$; the
mean cluster size $S(\phi)= \sum s^2 n(s)\sim
(\phi_c-\phi)^{-\gamma}$, with $\gamma=1.8\pm0.1$; the
connectedness length $\xi\sim(\phi_c-\phi)^{-\nu}$, with
$\nu=0.88\pm0.01$; and the fractal dimension of large clusters
is $D_f=2.4\pm0.1$.
In the following we fix the number of particles, $N=1000$, where the threshold
is $\phi_c \simeq 0.095$.

\begin{figure}
\begin{center}
\includegraphics[width=7cm]{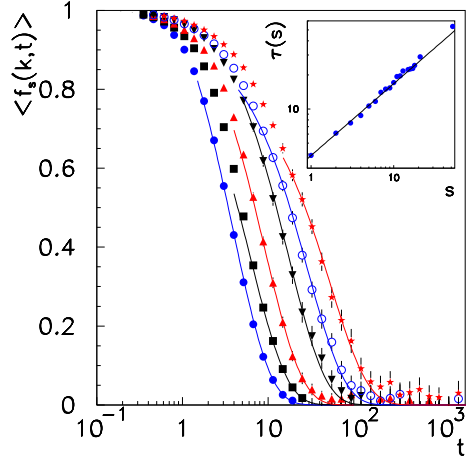} 
\end{center}
\caption{(Color online) {\bf Main frame}:
The self ISF, $\lan f_s(k,t)\ran$ for clusters of size
$s=1,2,4,10,21,52$, 
for $\phi=0.09$, 
$k=k_{min}$.
The curves are exponential
functions, $e^{-t/\tau(s)}$. {\bf Inset}: The relaxation time,
$\tau(s)$, as function of the cluster size $s$ for $k=k_{min}$ and
$\phi=0.09$. The continuous curve is a power law $s^{x}$ with
$x\simeq 0.65$.} \label{fig2}
\end{figure}

\subsection{Size dependence of dynamical behaviour of the clusters}

In the sol phase we have studied the dynamical behaviour of the clusters as a 
function of the size $s$. 
In particular we have measured the self ISF and the mean squared displacement 
of clusters, respectively 
$\lan f_s(k,t)\ran\equiv\frac{1}{s}\overline{\sum_{i\in C_s}
\lan e^{i\vec{k}\cdot(\vec{r}_i(t)-\vec{r}_i(0))}\ran}$, and  
$\lan \Delta r^2(s,t)\ran\equiv \frac{1}{s}\overline{\sum_{i\in C_s}\langle
|\vec{r}_i(t)- \vec{r}_i(0)|^2\rangle}$, where again $\overline{\dots^{}}$ is
the average of all clusters $C_s$ with fixed size $s$.

After an initial transient, we find that $\lan f_s(k,t)\ran$ for $k=k_{min}$ 
is well fitted by exponential tail, $e^{-t/\tau(s)}$ (Fig.\ref{fig2}), 
with $1/\tau(s) \propto s^{-x}$ (Inset of Fig.\ref{fig2}) 
and $x$  not depending on the volume fraction.
Our data furnishes\cite{nota_tau}  $x\sim 0.65$.

\begin{figure}
\begin{center}
\includegraphics[width=7cm]{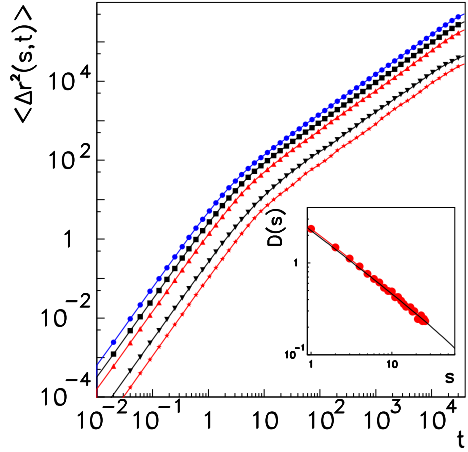} 
\end{center}
\caption{(Color online) {\bf Main frame}: The mean squared displacement, $\lan
\Delta r^2(s,t)\ran$, for $\phi=0.09$, and clusters of size
 $s=1,2,4,10,21,52$. {\bf Inset}: The diffusion coefficient,
$D(s)$, as function of the cluster size $s$ for $\phi=0.09$. The continuous 
curve is a power law $s^{-x}$ with
$x\sim 0.67$.} 
\label{fig2_bis}
\end{figure}

The mean squared displacement of clusters (shown in the main frame of 
Fig. \ref{fig2_bis}), after a ballistic regime at short time, displays a 
diffusive behaviour. The diffusion coefficient of clusters, $D(s)$, obtained 
as $\lan \Delta r^2(s,t)\ran\simeq 6 D(s)t $, is plotted
in the inset of Fig. \ref{fig2_bis} as a function of size $s$. 
In agreement with the results for the relaxation time, 
we find that $D(s) \propto s^{-x}$, with $x\simeq 0.67$.

\subsection{Self ISF and its fluctuations}

\begin{figure}
\begin{center}
\includegraphics[width=7cm]{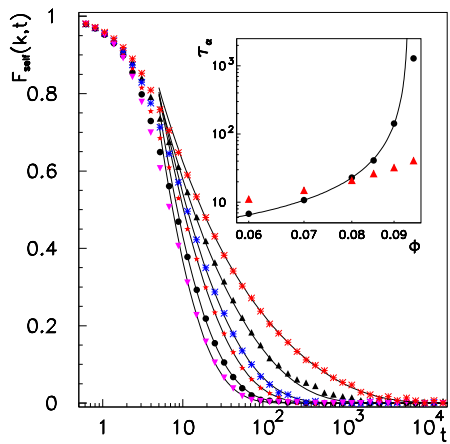} 
\caption{(Color online) {\bf Main Frame}:
Self ISF, $F_{self}(k,t)$ for $k=k_{min}$ and $\phi=0.06$, $0.07$,
$0.08$, $0.085$, $0.09$, $0.095$ (from left to right) as a function of
time $t$. The lines are fitting curves: $At^{-0.36}e^{-(t/\tau_\alpha)^{0.6}}$.
{\bf Inset}:
Structural relaxation time, $\tau_\alpha$ (full circles), 
compared with 
the inverse of the diffusion coefficient $20/D$ 
(full triangles) as a function of
the volume fraction.
The full line is the fitting curve: $\tau(\phi) \propto (\phi_c-\phi)^{-f}$,
with $f \simeq 1.4$ and $\phi_c\simeq 0.095$.
}
\label{fig_self}
\end{center}
\end{figure}

In the sol phase, due to the superposition of
the contributions of different clusters, the self ISF is expected to 
follow Eq.(\ref{conto4}):
\begin{equation}
F_{self}(k_{min},t)\propto~t^{-c_1} e^{-(t/\tau_\alpha)^\beta},
\end{equation}
where $\beta=1/(x+1)\simeq 0.60$, $c_1=\beta(\tau-3/2)\simeq 0.36$, and
$\tau_\alpha\propto~(\phi_c-\phi)^{-f}$, with $f=x D_f \nu\simeq 1.4$. 
$F_{self}(k_{min},t)$ is plotted in Fig. \ref{fig_self}
for $\phi<\phi_c$. After the initial transient, the data are well fitted by 
the function, Eq.(\ref{conto4}) 
(continuous curves in figures). 
Furthermore, in agreement with theoretical predictions, the relaxation time 
$\tau_\alpha$ (plotted in the Inset of
Fig.\ref{fig_self} as a function of $\phi$)
appears to diverge approaching the transition threshold 
with the exponent $f\simeq 1.4$.
At the threshold $\phi_c\simeq 0.095$ finite size effects appear.

Interestingly the presence of dynamical heterogeneities,
i.e. groups of particles with different diffusion coefficient, is related to the
breakdown of Stokes-Einstein relation.
As shown in \ref{appendix1}, the relaxation time $\tau_\alpha$
is essentially the relaxation time of the critical cluster.
On the contrary the diffusion coefficient of the system,
obtained as $\lan \Delta r^2(t)\ran\simeq 6 Dt $,  is given by the average over
clusters with different sizes, $D\equiv \overline{D}=\sum_s s n(s) D(s)$.
Since $D(s)\propto s^{-x}$, it is clear that
$D$ is dominated by small clusters. As a consequence,
although $\tau_\alpha$ diverges at the threshold,
$D$  does not go to zero
at $\phi_c$ (see Inset of Fig.\ref{fig_self}),
due to the diffusion of
small clusters through the gel matrix even for $\phi \gtrsim \phi_c$.

\begin{figure}
\begin{center}
\includegraphics[width=7cm]{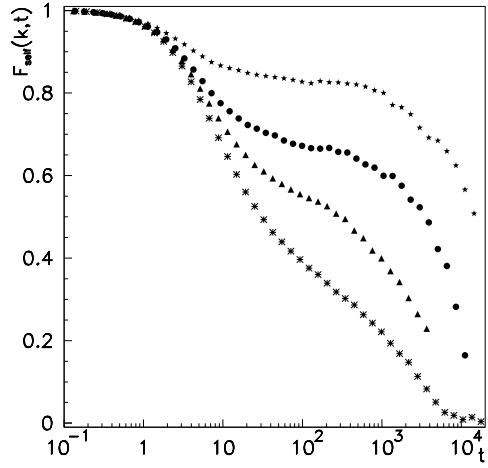} 
\caption{(Color online) Self ISF,
$F_{self}(k,t)$ for $k=k_{min}$ and $\phi=$
$0.1$,$0.105$,$0.11$,$0.12$.
}
\label{fig_gel}
\end{center}
\end{figure}

\begin{figure}
\begin{center}
\includegraphics[width=7cm]{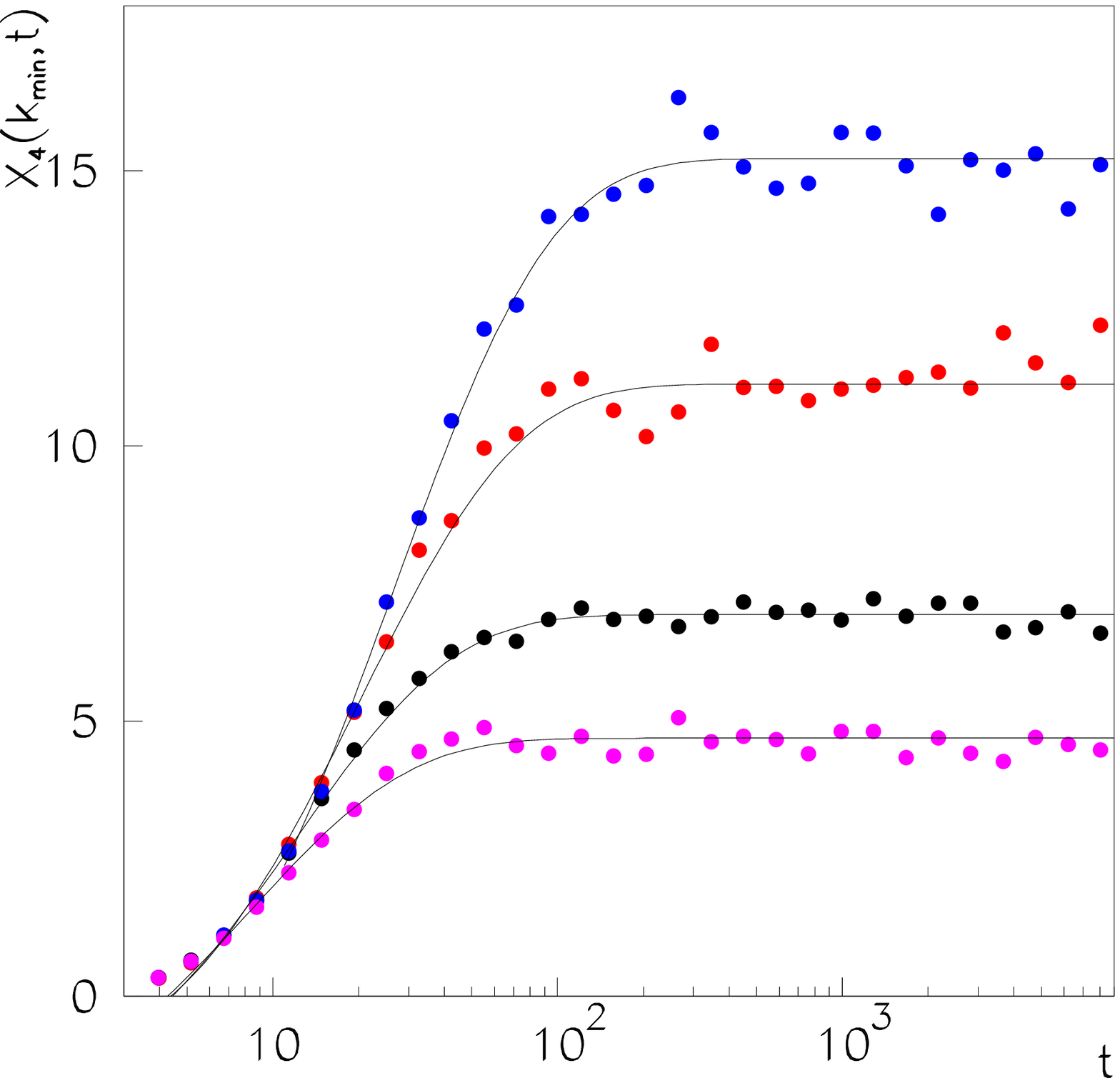} 
\caption{(Color online)
Dynamical susceptibility, $\chi_{4}(k_{min},t)$ for
$\phi=0.06$, $0.07$, $0.08$, $0.085$ (from bottom to top) as a function of
time $t$. The
lines are fitting curves: $S\cdot\left(1-At^{0.24}e^{-(t/\tau'_\alpha)^{0.6}}
\right)$.}
\label{fig_chi}
\end{center}
\end{figure}

In the gel phase a detailed analysis is not 
possible: $F_{self}(k_{min},t)$   (plotted in Fig.\ref{fig_gel} for 
$\phi>\phi_c$) displays a plateau, however at long times, 
due to finite size effects,
it relaxes to zero following an exponential 
function.

Finally in the sol phase,
we have also measured the dynamical susceptibility, $\chi_4(k,t)$,
defined as the fluctuations of the self ISF, given by Eq.(\ref{chi4}).
In Fig.\ref{fig_chi},
$\chi_4(k,t)$ is plotted for $k=k_{min}$ and different volume
fractions. After the initial transient, the approach to the plateau is well
fitted by Eq.(\ref{conto6}):
\begin{equation}
\lim_{k\to 0}\chi_4(k,t)\simeq~S\cdot\left(1-At^{c_2} e^{-(t/\tau'_\alpha
)^\beta}\right),
\end{equation}
where $\beta=1/(x+1)\simeq 0.6 $ and $c_2=\beta(5/2-\tau)\simeq 0.24$.
Note that the relaxation time $\tau'_\alpha$ coincides with $\tau_\alpha/2$
only at low volume fraction; near the threshold $\tau'_\alpha$ is instead lower
than  $\tau_\alpha/2$, due to
the contribution of disconnected particles at intermediate times.

\begin{figure}
\begin{center}
\includegraphics[width=7cm]{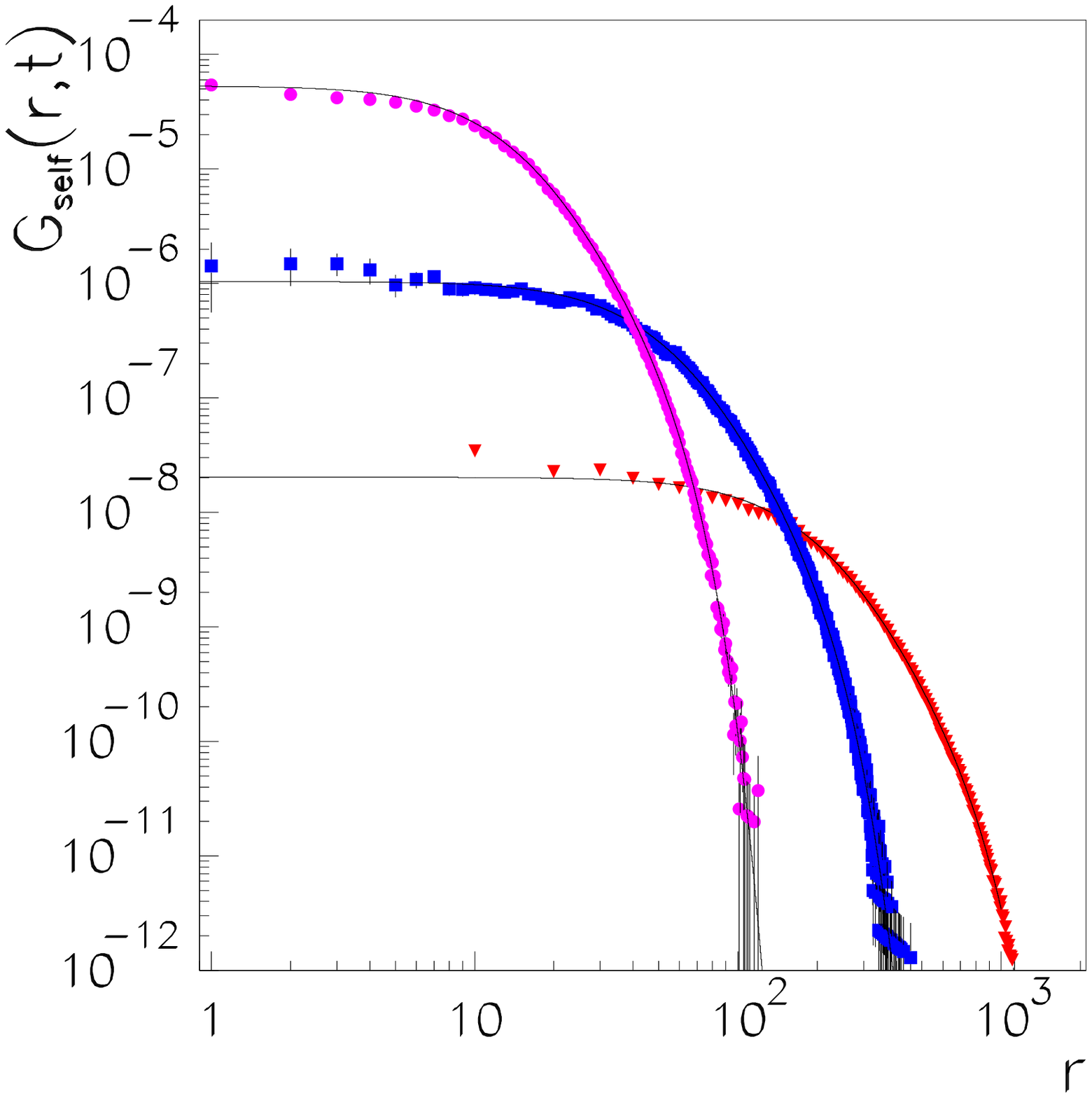}
\caption{(Color online)
The self part of the Van-Hove
distribution for $\phi=0.07$ and time $t=~93.199,~1285.02, ~17715.2$
(from top to bottom). Full lines are
obtained from Eq.(\ref{self-van-hove}).} \label{fig:gr}
\end{center}
\end{figure}

\subsection{Self part of the Van-Hove function and the non-Gaussian parameter}

\begin{figure}
\begin{center}
\includegraphics[width=7cm]{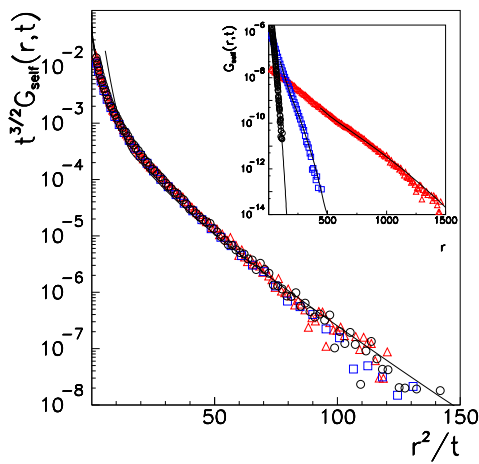} 
\caption{(Color online) {\bf Main frame}:
$t^{3/2}G_{self}(r,t)$ as a function of $r^2/t$ for $\phi=0.07$ and
$t=~93.199, ~1285.02, ~17715.2$. The full line is obtained from 
Eq.(\ref{conto:fitg}).
{\bf Inset}: The self part of the Van-Hove
distribution for $\phi=0.07$ and time $t=~93.199,~1285.02, ~17715.2$
(from top to bottom). Full lines are obtained 
from Eq.(\ref{conto:fitg}) 
with the values of $s^*$, $x$, $\tau$ and $a$ measured in the simulations.
Note that the curves shown in figure are numerically indistinguishable from 
exponential functions.} \label{fig_fitg}
\end{center}
\end{figure}

In the sol phase we have also measured the self part of the Van-Hove function,
$G_{self}(r,t)$, defined by Eq.(\ref{gself0}).
In the long time regime, $G_{self}(r,t)$ 
is fitted by a Gaussian curve only for short distances, and 
it seems well fitted by an exponential function for long distances 
\cite{tiziana_pre}.  In the long time regime, 
clusters of
any size show a diffusive behaviour (see Fig. \ref{fig2_bis}), then
we have suggested \cite{tiziana_pre} that $G_{self}(r,t)$ is given by a 
superposition of Gaussians, Eq.(\ref{self-van-hove}):
\begin{equation}
G_{self}(r,t)=\left[\sum_s \frac{s n(s)}{(4\pi
D(s)t)^{3/2}}\exp\left(-\frac{r^2}{4D(s)t}\right)\right],
\label{eq24}
\end{equation}
where $D(s)$ is the diffusion coefficient of clusters of size $s$, and $n(s)$ 
is the cluster size distribution.
Data well agree with our hypothesis, as indicated in Fig.\ref{fig:gr},
where we have used $n(s)$ and $D(s)$ measured in the simulations.
As shown in the main frame of Fig. \ref{fig_fitg}, $t^{3/2}G_{self}(r,t)$
plotted as a function of $r^2/t$ for fixed $\phi$ and for different times after the
initial transient,
collapse onto a single master curve,  
supporting the
hypothesis that the data satisfy Eq.(\ref{eq24}). 

Finally the comparison with the approximate form obtained in \ref{appendix3} for
$r^2\gg t$, Eq. (\ref{eq17}), 
\begin{equation}
t^{3/2}G_{self}(r,t)\propto
\left(-A+\frac{1}{s^*}+\frac{xr^2}{4at}\right)^{-1}
\exp\left[-\frac{r^2}{4at}\right],
\label{conto:fitg}
\end{equation}
(with $A\equiv 1-\tau+3x/2$, $s^*$ and $a$ obtained from the simulations)
gives again a good agreement for high enough $r$ (Fig.\ref{fig_fitg} and Inset).
Note that the curves obtained from Eq.(\ref{conto:fitg})
and shown in figure are numerically indistinguishable from exponential 
functions in the considered range
\cite{nota_exp} (Inset of Fig.\ref{fig_fitg}).

\begin{figure}
\begin{center}
\includegraphics[width=7cm]{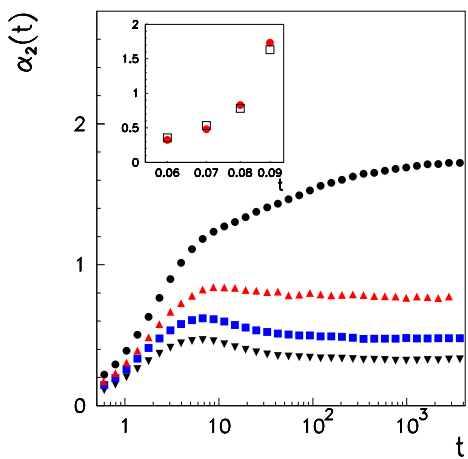} 
\caption{(Color online) {\bf Main frame}: Non-Gaussian parameter,
$\alpha_2(t)$, as a function of time $t$ for $\phi=~0.06$, $0.07$, $0.08$,
$0.09$ (from bottom to top). 
{\bf Inset} Asymptotic value of $\alpha_2(t)$ (empty squares) compared with the 
fluctuations of the diffusion coefficient given by  Eq.(\ref{fluttD}) (full
circles).  } \label{alfa}
\end{center}
\end{figure}

In agreement with the above picture, we expect that the non-Gaussian 
parameter tends in the long time limit to a plateau, whose value coincides 
with the fluctuations of the diffusion coefficient, Eq.(\ref{fluttD}).
In the main frame of Fig.\ref{alfa} the non-Gaussian parameter is plotted for
different volume fractions, and in the inset of Fig.\ref{alfa}  the plateau
value is compared with the
fluctuations of the diffusion coefficient. The data are in good agreement with
Eq.(\ref{fluttD}), confirming that
in permanent gels the non-gaussianity of the displacement distribution
is due to the superposition of the contributions of clusters of 
different sizes.
It is worth to notice that the main contribution to $\alpha_2(t)$ comes
from small finite clusters. In fact, the bigger the cluster,
the lower its diffusion coefficient and hence its contribution 
to the non-Gaussian parameter. Therefore, no criticality
of the plateau value of the non-Gaussian parameter
is observed approaching the transition threshold.

\section{Conclusions}
\label{sect4}

In this paper we show how the complex dynamics, such as stretched 
exponentials and power law behaviors,
observed experimentally and numerically in gelling systems, emerges
from the contribution of single clusters, which instead decay with a simple
exponential. 
Furthermore, we establish a connection between this complex 
behaviour and critical exponents of percolation theory.

We also find, in the diffusive regime, an asymptotic form
(for long enough distances) of the self part of
the Van-Hove function, which deviates from the Gaussian distribution, and is 
numerically very similar to
an exponential tail, usually observed in a large variety of complex system. 
Our finding suggest therefore that such deviation from Gaussianity  
may be due to a general mechanism, 
which may be ascribed to the presence of heterogeneities.

The theoretical predictions are found in agreement with numerical results, 
that we find in the FENE model for permanent gels.
We suggest that a similar analysis can be extended to
systems with finite lifetime bonds, as colloidal gels, glassy systems or spin
glasses, where a ``suitable'' definition of clusters is necessary.

\newpage
\appendix
\section{Self Intermediate Scattering Functions }
\label{appendix1}
In this section we show in details the calculations which gives the 
predictions shown in Sect. \ref{sect3} for the time dependence of the self ISF
in the thermodynamic limit.

Starting from Eq.(\ref{integral}), in the hypothesis that 
$f_s(k_{min},t)\simeq e^{-t/\tau(s)}$, and
$1/\tau(s)\sim a s^{-x}$, the self ISF for a given bond configuration becomes:
\begin{eqnarray}
F_{self}(k_{min},t)\simeq \int ds~s n(s) e^{-a t s^{-x}}.
\label{eq:A1}
\end{eqnarray}

Let us consider three different cases: (i) $\phi=\phi_c$; (ii) $\phi<\phi_c$;
(iii) $\phi>\phi_c$.

For $\phi\le\phi_c$, 
$n(s)$ can be
written as $n(s) \simeq s^{-\tau} \exp(-s/s^*)$ \cite{stauffer}, where $s^*$ is
a cutoff value \cite{nota_k}.
Then we obtain:
\begin{eqnarray}
F_{self}(k_{min},t)\simeq \int ds~\exp(t(s)),
\label{sol}
\end{eqnarray}
where 
\begin{equation}
t(s)\equiv-(\tau-1)\ln s-\frac{s}{s^*}-\frac{a t}{s^x}.
\label{ts}
\end{equation}

(i) At the gelation threshold, $\phi=\phi_c$, $s^*\to \infty$, and the
integral, Eq.(\ref{sol}), with $t(s)$ given by Eq.(\ref{ts}), can be
evaluated exactly:
\begin{eqnarray}
F_{self}(k_{min},t)\propto~t^{-z}\Gamma(z),
\end{eqnarray}
where $\Gamma(z)\equiv\int_0^\infty d\sigma~\sigma^{z-1}\exp(-\sigma)$ is the
$\Gamma$-function with $z=(\tau-2)/x$.

(ii) In the sol phase, $\phi<\phi_c$, we are able to give only approximated
predictions.
The function $t(s)$, given by Eq.(\ref{ts}),
has a maximum for $\tilde{s}$ such that \cite{nota_tilde}
\begin{equation}
\frac{\tilde{s}}{s^*}=-(\tau-1)+\frac{xat}{\tilde{s}^{x}}.
\label{tilde}
\end{equation}
Let us approximate $t(s)$ with
$t(\tilde{s})-(s-\tilde{s})^2/(2\sigma^2)$, where
\begin{eqnarray}
\frac{1}{\sigma^2}
\equiv -\left.\frac{d^2t(s)}{ds^2}\right|_{s=\tilde{s}}
=\frac{1}{\tilde{s}^2}\left(-(\tau-1)+x(x+1)\frac{at}{\tilde{s}^x}
\right).
\end{eqnarray}
If $\tilde{s}\gg\sigma$, $\int
ds~\exp\left[-\frac{(s-\tilde{s})^2}{2\sigma^2}\right]
= (2\pi\sigma^2)^{1/2}$, 
and
\begin{eqnarray}
F_{self}(k_{min},t)  \simeq \exp\left[t(\tilde{s})\right]\int
ds~\exp\left[-\frac{(s-\tilde{s})^2}{2\sigma^2}\right]
\nonumber \\
\propto
\frac{1}{{\tilde{s}}^{(\tau-2)}}\left(-(\tau-1)+x(x+1)\frac{at}{\tilde{s}^x}
\right)
^{-1/2}
\exp\left[-\left(\frac{\tilde{s}}{s^*}+\frac{at}{\tilde{s}^x}\right)\right].
\label{conto3}
\end{eqnarray}

Let us consider two limit cases: (1) $s^*\to\infty$; (2) 
$at/\tilde{s}^x\to\infty$.

(1) Using Eq.(\ref{tilde}), Eq.(\ref{conto3}) can be written 
in the following way:
\begin{eqnarray}
F_{self}(k_{min},t)\propto
\frac{\exp\left[-\frac{x+1}{x}\frac{\tilde{s}}{s^*}\right]
}{{\tilde{s}}^{(\tau-2)}}
\left(x(\tau-1)+(x+1)\frac{\tilde{s}}{s^*}\right)^{-1/2},
\end{eqnarray}
which in the limit $s^*\to\infty$, where $\tilde{s}\simeq 
\left(xat/(\tau-1)\right)^{1/x}$, gives 
again $F_{self}(k_{min},t)\propto~t^{-z}$ with
$z=(\tau-2)/x$, in agreement with previous calculations.

(2) Using Eq.(\ref{tilde}), Eq.(\ref{conto3}) can be written in the following 
way:
\begin{eqnarray}
F_{self}(k_{min},t)\propto
\frac{\exp\left[-\frac{(x+1)at}{\tilde{s}^x}\right]
}{{\tilde{s}}^{(\tau-2)}}
\left(-(\tau-1)+x(x+1)\frac{at}{\tilde{s}^x}
 \right)^{-1/2}.
\end{eqnarray}
In the limit $at/\tilde{s}^x \to\infty$, 
$\tilde{s}\simeq (x a s^* t)^{1/(x+1)}$, and we obtain
\begin{eqnarray}
F_{self}(k_{min},t)\propto~t^{-c_1} e^{-(t/\tau_\alpha)^\beta},
\end{eqnarray}
where $\beta=1/(x+1)$, $c_1=\beta(\tau-3/2)$, and
$\tau_\alpha\propto~{s^*}^{x}$, which diverges at the threshold as
power law with the exponent $f=x D_f \nu$. 

(iii) Finally in the gel phase, $\phi>\phi_c$,
$sn(s)$ in Eq.(\ref{eq:A1})
can be written as \cite{nota_delta, stauffer}
$sn(s)=P_\infty\delta_{s,s_{max}}+Cs^{-\tau+1}e^{-(s/s^*)^{(d-1)/d}}$,
where $P_\infty$ is fraction of particles belonging to
the percolating cluster, 
$s_{max}$ is infinite in the thermodynamic limit, 
$d$ is the spatial dimension, and $C$ is a constant. 
Then, from Eq. (\ref{eq:A1}), we obtain:
\begin{eqnarray}
F_{self}(k_{min},t)\simeq P_\infty+C\int ds~\exp(t(s)),
\label{gel}
\end{eqnarray}
where
\begin{equation}
t(s)\equiv-(\tau-1)\ln s-\left(\frac{s}{s^*}\right)^{(d-1)/d}-\frac{a t}{s^x}.
\label{ts_gel}
\end{equation}

Following the same arguments as in the sol phase, 
the second term in Eq. (\ref{gel}) is written as:
\begin{eqnarray}
\int ds~\exp(t(s)) 
&\propto& \tilde{s}^{-(\tau-2)}\exp\left[-\left(1+\frac{xd}{d-1}
\right)\left(\frac{at}
{\tilde{s}^x}\right)\right] 
\nonumber \\
&\cdot& \left(-\frac{(\tau-1)(d-1)}{d}
+x\left(x+1-\frac{1}{d}\right)\frac{at}{\tilde{s}^x}
\right)^{-1/2},
\label{contogel_new}
\end{eqnarray}
where the maximum point, $\tilde{s}$, of $t(s)$, Eq. (\ref{ts_gel}),
is obtained from
\begin{equation}
-(\tau-1)+\frac{axt}{\tilde{s}^x}=
\frac{d-1}{d}\left(\frac{\tilde{s}}{s^*}\right)^{(d-1)/d}.
\end{equation}

In the limit $at/\tilde{s}^x \to\infty$,
$\tilde{s}\propto t ^{1/(x+1-1/d)}$, and we obtain
\begin{eqnarray}
F_{self}(k_{min},t)\simeq P_\infty+At^{-g} e^{-(t/\tau_\alpha)^{\beta_g}},
\end{eqnarray}
where $\beta_g=(1-1/d)/(1-1/d+x)$, $g=\beta_g (\tau-3/2-1/2d)/(1-1/d)$.

For a finite system however $s_{max}$ is finite, and the behavior of
$F_{self}(k_{min},t)$ is given by:
\begin{equation}
F_{self}(k_{min},t)\simeq \left\{ \begin{array}{ll} 
P_\infty+At^{-g} e^{-(t/\tau_\alpha)^{\beta_g}}
 &t\ll\tau_{max} \\
P_\infty~ e^{-t/\tau_{max}} &t>\tau_{max},  \end{array} \right.
\label{eq:A_finite}
\end{equation}
where 
$\tau_{max}=\tau(s_{max})$ is the relaxation time of the 
percolating cluster.

\section{Fluctuations of the self  ISF}
\label{appendix2} 

In the hypothesis of the previous section 
(i.e. $\lan f_s(k,t)\ran\simeq e^{-t/\tau(s)}$ and
$1/\tau(s)\simeq a s^{-x}$) 
Eq.(\ref{eq16}) becomes \be
\lim_{k\to 0}\chi_4(k,t)=\sum_s s^2 n(s) (1-e^{-2t/\tau(s)}), \ee 
which, for $\phi<\phi_c$, by replacing the sum with the integral, gives:
\be \lim_{k\to
0}\chi_4(k,t)\simeq S-\int ds \exp(w(s)), \label{integral2}
\ee where 
\be 
w(s)\equiv -(\tau-2)\ln s-\frac{s}{s^*}-\frac{2 a t}{s^x}. \label{ts2}
\ee

Note that the function $w(s)$ 
has a maximum for $\overline{s}$ such that
\begin{equation}
\frac{\overline{s}}{s^*}=-(\tau-2)+\frac{2xat}{\overline{s}^{x}}.
\label{overline}
\end{equation}
Let us approximate $w(s)$ with
$w(\overline{s})-(s-\overline{s})^2/(2\sigma^2)$, where
\begin{eqnarray}
\frac{1}{\sigma^2}
\equiv -\left.\frac{d^2w(s)}{ds^2}\right|_{s=\overline{s}}=
\frac{1}{\overline{s}^2}\left( (\tau-2) -x(x+1) \frac{2at}{\overline{s}^x} 
\right).
\end{eqnarray}
If $\overline{s}\gg\sigma$, $\int
ds~\exp\left[-\frac{(s-\overline{s})^2}{2\sigma^2}\right]= 
(2\pi\sigma^2)^{1/2}$, and
\begin{eqnarray}
S-\lim_{k\to 0}\chi_4(k,t) \simeq 
\exp\left[w(\overline{s})\right]\int
ds\exp\left(-\frac{(s-\overline{s})^2}{2\sigma^2}\right)
\nonumber \\
\nonumber \\ \propto\frac{1}{\overline{s}^{(\tau-3)}}
\left((\tau-2)-x(x+1)\frac{2at}{\overline{s}^x}\right)^{-1/2}
\exp\left[-\left(\frac{\overline{s}}{s^*}+
\frac{2at}{\overline{s}^x}\right)\right]. \label{conto5}
\end{eqnarray}

The limit $at/\overline{s}^x\to\infty$,  where $\overline{s}\propto (x s^* t)^{1/(x+1)}$, gives
\begin{eqnarray}
\lim_{k\to 0}\chi_4(k,t)\simeq~S\cdot(1-At^{c_2} e^{-(2
t/\tau_\alpha)^\beta}),
\end{eqnarray}
where $\beta=1/(x+1)$ and $c_2=\beta(5/2-\tau)$.
The exponent $\beta$ is exactly the same which appears in Eq.(\ref{conto4}) 
for the decay
to zero of the self ISF. Note that the relaxation time in the stretched exponential
function is given by $\tau_\alpha/2$.

\section{Self part of the Van-Hove function}
\label{appendix3}
In the sol phase, where
after an initial transient the system is found in a diffusive regime,
we assume the validity of Eq.(\ref{self-van-hove}) for the self part
of the Van-Hove function. 
By replacing the sum with the integral,
and putting $D(s) \simeq a s^{-x}$ ,
$G_{self}(r,t)$ can be written as
\be
G_{self}(r,t)\simeq \frac{1}{(4\pi t)^{3/2}} \int ds~
\exp\left(z(s)\right),
\label{gself1}
\ee
where
\be
z(s)\equiv \left(1-\tau+\frac{3 x}{2}\right)\ln s-\frac{s}{s^*}-\frac{s^x r^2}{4 a t}.
\label{zs}
\ee
The condition for the first derivative of $z(s)$ to be zero gives:
\begin{equation}
\frac{s_m}{s^*}+\frac{xr^2 s^x_m}{4at}=1-\tau+\frac{3}{2}x.
\label{sm}
\end{equation}

This equation admits a solution $s_m>0$
only if $A\equiv1-\tau+3x/2>0$.
Under this condition the solution of Eq.(\ref{sm}) is a maximum, in fact
\begin{eqnarray}
-\left.\frac{d^2z(s)}{ds^2}\right|_{s=s_m}
&=&\frac{1}{s_m^2}\left(A+x(x-1)\frac{r^2 s^x_m}{4at}\right)>0,
\end{eqnarray}
is always true for $A>0$ (i.e. $x>2(\tau-1)/3$) \cite{notaA}.
In this case we can approximate $z(s)$ with
$z(s_m)-(s-s_m)^2/(2\sigma^2)$, where
$\frac{1}{\sigma^2}\equiv -\left.\frac{d^2z(s)}{ds^2}\right|_{s=s_m}$.
If $s_m\gg\sigma$, $\int
ds~\exp\left[-\frac{(s-s_m)^2}{2\sigma^2}\right]= (2\pi\sigma^2)^{1/2},$
and we can write:
\begin{eqnarray}
t^{3/2}G_{self}(r,t)  \propto \exp\left[z(s_m)\right]
\int ds~\exp\left[-\frac{(s-s_m)^2}{2\sigma^2}\right]
\nonumber \\
\propto
s_m^{1+A} \exp\left[-\frac{s_m}{s^*}-\frac{r^2 s^x_m}{4at}\right]
\left(A+x(x-1)\frac{r^2 s^x_m}{4at}\right)^{-1/2}.
\label{gself}
\end{eqnarray}

Let us consider two limit cases: (i) $r^2\ll at$, and (ii) $r^2\gg at$. From 
Eq.(\ref{sm}) we obtain:
\begin{equation}
s_m \simeq \left\{ \begin{array}{ll} A s^{*} &r^2\ll at \\
\left(\frac{4Aat}{xr^2}\right)^{1/x} &r^2\gg at,  \end{array} \right.
\end{equation}
and Eq.(\ref{gself}) becomes:
\begin{equation}
t^{3/2}G_{self}(r,t)\propto \left\{ \begin{array}{ll} 
\left(A+\frac{x(x-1)r^2 }{4a't}\right)^{-1/2}
\exp\left[-\frac{r^2}{4a't}
\right] &r^2\ll at \\
\left(\frac{4Aat}{xr^2}\right)^{(A+1)/x} \exp \left[-\frac{A'}{s^*}\left(
\frac{4at}{r^2}\right)^{1/x}\right] &r^2\gg at,  \end{array} \right.
\end{equation}
where $a'\equiv a/(As^*)^x$, and $A'=(A/x)^{(1/x)}$. Note that the condition
$s_m\gg \sigma$ is satisfied in the limit $r^2/ at\to 0$ where $s_m\to A s^{*}$,
which increases with increasing $x$ and/or approaching the gelation threshold.
In the opposite limit, $r^2/ at\to \infty$, $s_m\to 0$, and hence the condition 
$s_m\gg \sigma$ is not satisfied. In this case this approximation is 
expected not to hold and a development at the first order of the Taylor series 
of $z(s)$ around $s=1$ should be more appropriate (see below). 

In the case $A\equiv 1-\tau+3x/2<0$ (as in the FENE model for permanent gels 
presented in Sect. \ref{sect3}), $z(s)$ is a monotonic decreasing function of 
$s$:
\be
\frac{dz}{ds}=\frac{A}{s}-\frac{1}{s^*}-\frac{xr^2 s^{x-1}}{4at}<0.
\ee
Let us develop $z(s)$ at the first order of the Taylor series around the maximum
$s_m=1$:
\be
z(s)\simeq z(s_m)+\left.\frac{dz}{ds}\right|_{s=s_m}(s-s_m).
\label{zs1}
\ee 
By replacing Eq.(\ref{zs1})  
in Eq.(\ref{gself1}), we obtain
\begin{eqnarray}
t^{3/2} G_{self}(r,t) \propto 
\left(-\left.\frac{dz}{ds}\right|_{s=s_m}\right)^{-1}
\exp\left[z(s_m)\right]\nonumber\\
\propto \left(-A+\frac{1}{s^*}+\frac{xr^2}{4at}\right)^{-1}
\exp\left[-\frac{r^2}{4at}\right].
\end{eqnarray}
This approximation is expected to hold in
the limit of long distances ($r^2\gg at$), where
the second order of the Taylor series 
around $s_m=1$ is much smaller than the first one, and only a small interval of 
values of $s$ around $s_m=1$ contribute to the integral in Eq.(\ref{gself1}).

\hspace{1cm}

The research is supported by CNR-INFM Parallel Computing Initiative,
S.Co.P.E. and L.R. N.5 2005.


\begin{thebibliography}{99}
\bibitem{nota_tau} As the relaxation time $\tau$ is dominated by the
longest time $\tau(s^*)$, we assume 
$\tau \simeq \tau(s^*) \propto {s^*}^{x}$. 
Using that \cite{stauffer}  $s^* \propto\xi^{D_f}$, 
$\tau \simeq \xi^{2-\beta/\nu}$ and $2-\beta/\nu=D_f -1$, 
we obtain $x=(D_f-1)/D_f$, which in $d=3$ random percolation gives 
$x \simeq 0.6$.
\bibitem{nota_exp} The function $g(r,t)=\ln[G_{self}(r,t)]$ can be
approximated by a Taylor expansion
around $r=r_0$. We observe that for $r\rightarrow \infty$,
the second order term of the
expansion of $g(r,t)$ is negligible respect the first order one,
indicating that the curve in Eq.(\ref{conto:fitg})
resembles an exponential function of $r$ in the asymptotic regime.
For finite $r$ it is always possible to find a range,
whose amplitude depends on $r_0$ and $t$,
in which the first order expansion represents a good approximation.
\bibitem{nota_k} It is worth to notice that the cutoff value $s^*$
depends on the wave vector $k$. In fact $s^*=\xi^{D_f} f(k\xi)$, where
$f(z)$ is a function which tends to a constant for small $z$,
whereas it behaves as $z^{-D_f}$ for large values of $z$.
\bibitem{nota_tilde}
If $x=1$, the equation, Eq.(\ref{tilde}), for
$\tilde{s}$ becomes $\tilde{s}^2/s^*+(\tau-1)\tilde{s}-t=0$, which admits
only one solution for $\tilde{s}>0$, $\tilde{s}=s^*\left(-(\tau-1)+
\sqrt{(\tau-1)^2+4t/s^*}\right)/2$.
\bibitem{nota_delta} Actually we should consider
$sn(s)=P_\infty g_{s,s_{max}}+Cs^{-\tau+1}e^{-(s/s^*)^{(d-1)/d}}$, 
where $g_{s,s_{max}}$ is
a distribution peaked around  $s_{max}$, which tends to $\delta_{s,s_{max}}$
for $N\to \infty$ and $s_{max}\to \infty$.
\bibitem{notaA}
By writing $\left.\frac{d^2z(s)}{ds^2}\right|_{s=s_m}$ in the following way:
\begin{eqnarray}
\left.\frac{d^2z(s)}{ds^2}\right|_{s=s_m}
&=&-\frac{1}{s_m^2}\left(x A+(1-x)\frac{s_m}{s^*}\right)=\nonumber \\
&=&-\frac{1}{s_m^2}\left(A+x(x-1)\frac{r^2 s^x_m}{4at}\right),
\end{eqnarray}
we easily see that, for $A>0$, $\left.\frac{d^2z(s)}{ds^2}\right|_{s=s_m}$ 
is always less
than zero, for both $x>1$ and $x<1$.
\end{thebibliography}
\end{document}